# "Operating Spacecraft Around Comets:

# Evaluation of the Near-Nucleus Environment"

**C.M. Lisse[1], M.R. Combi[2], T.L. Farnham[3], N. Dello Russo[1], S. Sandford[4], A.F. Cheng[1], U. Fink[5], W.M. Harris[5], J. McMahon[6], D.J. Scheeres[6], H.A. Weaver[1], J. Leary[1]**

[1] Johns Hopkins University Applied Physics Laboratory, Laurel, MD 20723   carey.lisse@jhuapl.edu, andy.cheng@jhuapl.edu, neil.dello.russo@jhuapl.edu, hal.weaver@jhuapl.edu, james.leary@jhuapl.edu

[2] University of Michigan, Ann Arbor, Michigan 48109  mcombi@umich.edu

[3] Department of Astronomy, University of Maryland College Park, College Park, MD    farnham@astro.umd.edu

[4] Astrophysics Branch, Space Sciences Division, NASA/Ames Research Center, Moffett Field, CA, USA 94035 scott.a.sandford@nasa.gov

[5] Lunar & Planetary Laboratory, University of Arizona, 1629 E. University Blvd Tucson AZ 85721  wharris@lpl.arizona.edu, uwefink@lpl.arizona.edu

[6] Department of Aerospace Engineering Sciences, Colorado Center for Astrodynamics Research, University of Colorado, Boulder, CO 80305    scheeres@colorado.edu, jay.mcmahon@colorado.edu

**38 Pages, 15 Figures, 1 Table**

Proposed Running Title: **"Operating Spacecraft in the Near-Nucleus Cometary Environment"**

Please address all future correspondence, reviews, proofs, etc. to:

Dr. Carey M. Lisse

Planetary Exploration Group, Space Exploration Sector

Johns Hopkins University, Applied Physics Laboratory

SES/SRE, Building 200, E206

11100 Johns Hopkins Rd

Laurel, MD 20723

240-228-0535 (office) / 240-228-8939 (fax)

Carey.Lisse@jhuapl.edu

# Abstract

We present a study of the current state of knowledge concerning spacecraft operations and potential hazards while operating near a comet nucleus. Starting from simple back of the envelope calculations comparing the cometary coma environment to benign conditions on Earth, we progress to sophisticated engineering models of spacecraft behavior, and then confront these models with recent spacecraft proximity operations experience (e.g., *Rosetta*). Finally, we make recommendations from lessons learned for future spacecraft missions that enter into orbit around a comet for long-term operations. All of these considerations indicate that, with a proper spacecraft design and operations planning, the near-nucleus environment can be a relatively safe region in which to operate, even for an active short period comet near perihelion with gas production rates as high as $10^{29}$ molecules/s. With gas densities similar to those found in good laboratory vacuums, dust densities similar to Class 100 cleanrooms, dust particle velocities of 10's of m/s, and microgravity forces that permit slow and deliberate operations, the conditions around a comet are generally more benign than a typical day on Mars. Even in strong dust jets near the nucleus' surface, dust densities tend to be only a few grains/cm$^3$, about the same as in a typical interior room on Earth. Stochastic forces on a modern spacecraft with tens of square meters of projected surface area can be accounted for using modern Attitude Control Systems to within tens of meters' navigation error; surface contamination issues are only important for spacecraft spending months to years within a few kilometers of the nucleus' surface; and the issues the Rosetta spacecraft faced, confusion of celestial star trackers by sunlit dust particles flying past the spacecraft, will be addressed using the next generation of star trackers implementing improved transient rejection algorithms.

# 1.    Introduction.

The purpose of this paper is to enumerate and study in detail the importance of physical processes acting on spacecraft operating in the near-nucleus comet environment, as an aid in the design of future comet rendezvous, landing, and sample return missions.

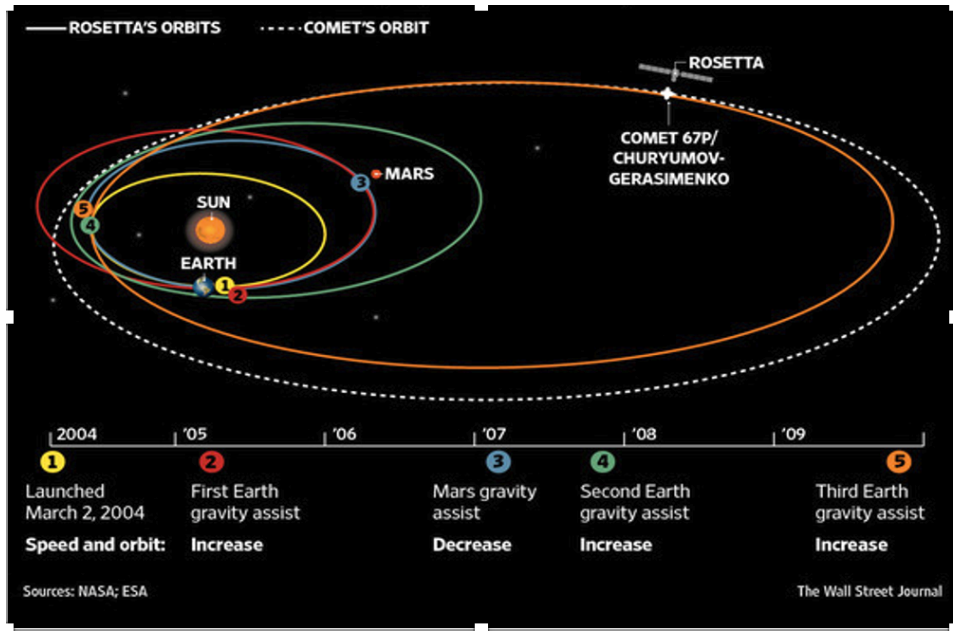

**Figure 1 - It took the Rosetta spacecraft more than 11.5 years and four planetary gravity assists to travel from the Earth to comet 67P/Churyumov-Gerasimenko.** The delivery of spacecraft resources to the near cometary nucleus environment is thus a difficult and expensive task requiring a high degree of optimization given known environmental operating conditions.  [Image courtesy of ESA, NASA, and the Wall Street Journal.]

Future spacecraft will fly into the near-nucleus regions of a comet's coma for a number of reasons. First and foremost, comets are thought to contain some of the most unaltered material left over from the beginning of the Solar System and are thus of great scientific interest as "fossils of Solar System formation". Comets also contain water, ices, and organic ingredients useful for resource utilization, be it for in-space applications or delivery to bodies for water/atmospheric supply. Understanding these early solar system fossils and harvesting these resources will require operating on or near the comet's nucleus and collecting and returning comet nucleus samples to Earth. Comets also represent potential impact hazards, which could require operating near or on the nucleus to divert a possible Earth-impacting trajectory. All of these spacecraft mission functions will require lengthy orbital maneuvers and multi-year trajectories (Fig. 1), so understanding the near-nucleus environment is critical for optimizing operations and the implementation of precious spacecraft resources like mass, power, fuel, shielding, etc.

For the purposes of this study, we define the near-nucleus comet environment as the region of space within which gas outflow from the comet is in the collisional, classical ideal gas regime (rather than the thin ballistic flow regime outside of this zone). This is the region where the various processes (outflow wind buffeting, dust particle collisions, gas drag, etc.) could have the largest effect on a spacecraft. While the exact radius of this zone depends on the comet's activity level and coma gas temperature, for a typical near-Earth comet with $Q_{gas} < 10^{29}$ mol/s (A'Hearn *et al.* 1995, Ootsubo *et al.* 2012, Reach *et al.* 2013, Bauer *et al.* 2015, Combi *et al.* 2019) it is within ~100 km from the nucleus surface. The gas and dust supplied to this region of space comes, via sublimation of ices trapped in the comet, typically from 3 sources (Fig. 2): a broad, low level, relatively uniform emission; more spatially localized "jets"; and occasional stochastic outbursts (Bruck Syal *et al.* 2013; Farnham *et al.* 2013; Gulkis *et al.* 2015; El-Maarry *et al.* 2017, 2019; Thomas *et al.* 2019).

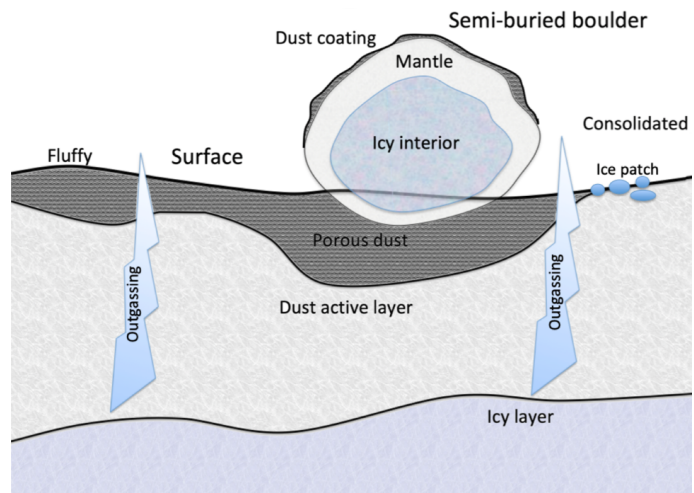

**Figure 2 – Schematic diagram showing the likely composition and structure of near-surface regions of a comet's nucleus.** Outgassing can come from diffuse sublimation of the deep icy layer through the overlying dust layers, or from small focused fractures or near-surface residual ice patches. (After Thomas *et al.* 2019.)

Some background on the present knowledge of the physical properties of comets is useful to the mechanics in play that can affect a spacecraft. Comets are ancient relic bodies of the solar system leftover from the very beginnings of the formation of our Solar System. Older than the Earth (Yu 2012, Golabek & Jutzi 2021), they consist of low density collections of the basic building blocks of the planets (estimated bulk porosity ~ 70 – 80%, estimated bulk density $\rho$ ~ 0.5 g/cm$^3$ for objects made of ~3:1 ratios of rocky dust with microcrystalline density of ~3.5 g/cm$^3$ and ice with microcrystalline density ~ 1g/cm$^3$; Lisse *et al.* 1998, 2004, 2006; Weiler *et al.* 2003; Gicquel *et al.* 2012; Davidsson *et al.* 2016; Choukroun *et al.* 2020). They are sourced in two large population collections in the modern Solar System. The first population resides in the Kuiper Belt (located at the edges of the original protoplanetary disk (PPD) surrounding the nascent Sun, 30 – 60 AU out) which follows the dynamical structure of the PPD. The second population resides in the Oort Cloud (located at 1000 – 100,000 AU from the Sun), a spherical shell of distant comets constructed as a result of building the giant planets. Oort cloud objects are in million year orbits that failed to merge with the nascent giants but were instead ejected into highly elliptical orbits with close to (but not quite equaling) escape velocity for the Solar System by close gravitational interactions with the giant planets. Both of these source populations exist in distant regions of the Solar System where icy phases like water and methanol are stable and act more like rock than the ice phases we know on Earth. However, when a body is perturbed out of these source regions and into the inner system (< 10 AU from the Sun), the comet's surface regions can become warm enough due to solar heating that these ices sublime and cause dust and gas to be shed into a temporary atmosphere, termed a "coma", that surrounds the comet nucleus. Since comets are small (radius < 30 km) and have low densities, gas emitted into the coma at the 100 – 400 K surface temperatures typical of active comets will not be gravitationally bound to the $\mu$g's of effective gravity at the surface of the comet source body (hereafter the "nucleus"). The surrounding coma structure is then produced by the balance of emitted cometary subliming volatiles + entrained dust at the comet's surface and through the collisional zone, and then the decoupling of the gas from the dust in the molecular flow region and the subsequent hydrodynamic outflow of the adiabatically expanding gases (Finson & Probstein 1968).

One critical, though typically unrecognized, aspect of spacecraft safety in the near-nucleus region is that operations can progress at a slow pace, minimizing hazards that arise from rapidly-changing conditions (i.e., cometary landings from a few km altitude take hours to conduct at m/s velocities, not the minutes at km/s velocities typical of planetary landings). The nature of cometary environments, with $\mu$g gravity and rarified atmospheres, means that physical processes tend to act slowly compared to what we are familiar with on Earth. First, as nuclei are only a few kilometers in size, they are low-mass objects with surface gravities $< 1$ cm/s$^2$ and orbital velocities that tend to be a few cm/s. Therefore, orbiting and maneuvering around the nucleus is done at a slow rate, on cadences of hours to days. Thus proximity operations permit maneuvers to be spaced many hours to days apart, and trajectory changes require only on the order of 1 cm/s delta-v. Descents and landings can be executed gradually, allowing ample time for ground-based operational control interacting with autonomous on-board operations to provide ample time for evaluating spacecraft circumstances and making decisions. If questionable conditions arise, a gentle thrust away from the surface will carry the spacecraft to a safe distance for re-evaluation of the situation.

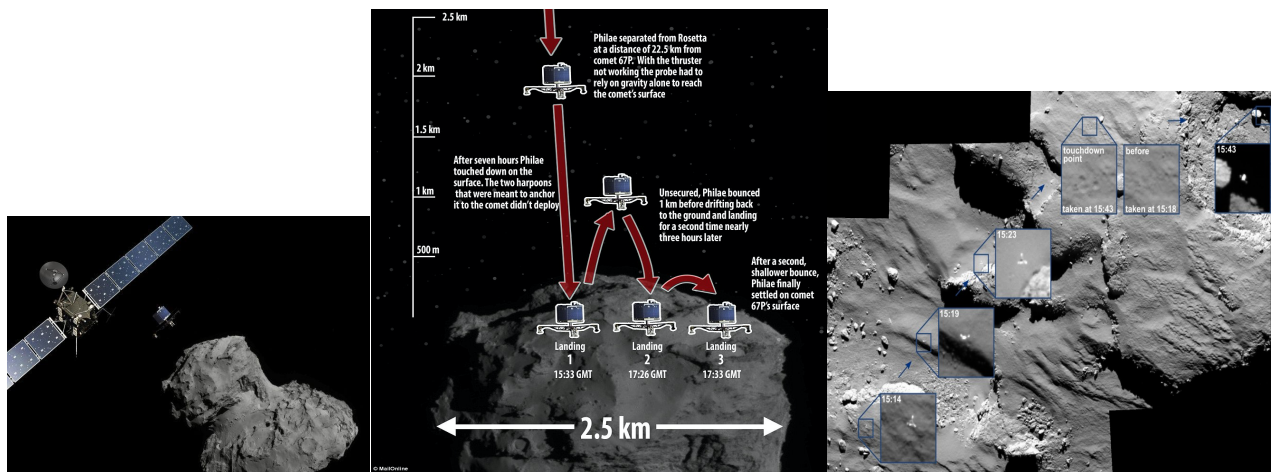

**Figure 3 – It took the Rosetta/Philae lander 7 hours to fall 22 km** from the mother spacecraft in 67P's microgee environment and hit its surface at a few m/s relative velocity, then another 2 hours to bounce off the surface and fall back to its final resting spot at 1-2 m/s. (*left*) Artist's impression of Philae falling from the Rosetta spacecraft bus to the nucleus. (*middle*) Cartoon schematic of the multi-touchdown path that Philae took upon encountering the nucleus. (*right*) Rosetta/Osiris NAC camera imagery of the lander along its nucleus ground track, showing the UT times of the snapshots and a flavor of the nucleus surface structures traversed during the ~1 km long ground track. (After Ulamec *et al.* 2016, 2017.)

In addition, a comet's general activity levels and mass outflow effects on a spacecraft evolve slowly, on timescales of weeks to months, as the comet travels towards/away from the Sun on its orbit, and on timescales of hours to days as the irregular nucleus rotates under the Sun's insolation (Marsden *et al.* 1973, Miller *et al.* 1989, Whipple 1989, Huebner *et al.* 1990). Occasionally, stochastic outbursts (thought to be due to landslides suddenly exposing fresh ices to solar heating) may introduce variations on timescales of hours, but these are rare and localized to geological faults and scarps (Steckloff *et al.* 2016, Vincent *et al.* 2016, Fink *et al.* 2021). The rate of material loss can vary from $10^5$ to $10^8$ kg/day for a nucleus of mass $10^{12} - 10^{14}$ kg with a $10^7$ to $10^9$ kg surrounding coma. (For comparison, consider that the mass of an Earth-like atmosphere extending out to 100 km from a 1 km radius body's surface would be ~1 x $10^{11}$ kg and that the mass of water in an average size terrestrial swimming pool is ~$1x10^5$ kg). Thus, comets are very small Solar System objects surrounded by highly rarified atmospheres.)

As with the attraction of gravity, the low mass loss rates mean low coma gas densities, and that repulsive forces due to outflowing gases also act on long timescales. Thus, spacecraft mission command, control, and operation (hereafter "conops") in the cometary environment can be performed slowly, allowing for low-thrust maneuvering supplied by small hydrazine jets or Solar-electric/Radionucleotide-Electric systems. A good example of these low speed, slowly developing operations was provided by the Rosetta mission's delivery of the Philae lander to the surface of comet 67P (Ulamec et al. 2016, 2017; Fig. 3).

### Table 1 - Comet Target Parameters for Past Spacecraft Missions[a]

| Comet | Mission | Mission Type | Closest Approach Distance(km) | Encounter Year(s) | Peri/Aphelion Distance (AU) | Max Outgassing Rate (mol/s) | Mean Nucleus Radius (km) |
|---|---|---|---|---|---|---|---|
| 1P/Halley | Giotto/Vega/ Suisei Sagdeev et al. 1986, Keller et al. 1986, Reinhard 1987 | Fast Flyby @ 68/78/80 km/s | 600/8000/1.5 x$10^5$ | 1985-1986 | 0.6 – 35 P = 75 yrs | 1 x $10^{29}$ | ~5.5 km |
| 26P/Grigg-Skjellerup | Giotto Grensemann & Schwehm 1993 | Fast Flyby @ 14 km/s | ~200 | 1992 | 1.2 - 4.9 P = 5.3 yrs | 7 x $10^{27}$ | ~1.3 km |
| 19P/Borrelly | DS-1 Rayman & Varghese 2001 | Fast Flyby @ 17 km/s | 2200 | 2001 | 1.4 - 5.8 P = 6.9 yrs | 3 x $10^{28}$ | ~2.4 km |
| 81P/Wild 2 | STARDUST Brownlee et al. 2014 | Fast Flyby @ 6.1 km/s / Coma Sample Return | 240 | 2004 | 1.6 - 5.3 P = 6.4 yrs | 1 x $10^{28}$ | ~2.1 km |
| 9P/Tempel1 | Deep Impact A'Hearn et al. 2005 | Fast Flyby / Impact @ 10 km/s | 0/500 | 2005 | 1.5 - 4.7 P = 5.6 yrs | 1 x $10^{28}$ | ~3.0 km |
| 103P/Hartley 2 | Deep Impact A'Hearn et al. 2011 | Fast Flyby @ 12 km/s | 700 | 2010 | 1.5 - 4.7 P = 5.6 yrs | 4 x $10^{28}$ | ~0.7 km |
| 9P/Tempel1 | STARDUST Veverka et al. 2013 | Fast Flyby @ 11 km/s | 180 | 2011 | 1.5 - 4.7 P = 5.6 yrs | 1 x $10^{28}$ | ~3.0 km |
| 67P/C-G | Rosetta Vallat, et al. 2017, Perez-Ayúcar et al. 2018 | Rendezvous & Landing | 0 | 2014-2018 | 1.2 - 5.7 P = 6.5 yrs | 2 x $10^{28}$ | ~1.6 km |

[a] – Comet data from Sosa & Fernandez 2009, ESA.

The initial comet reconnaissance of Giotto, Vega, Deep Space 1, Deep Impact, and Stardust missions (Keller et al. 1986; Reinhard 1988; Sagdeev et al. 1986; Rayman & Varghese 2001, Brownlee et al. 2004; A'Hearn et al. 2005, 2010; Veverka et al. 2013; Vallat, et al. 2017 Perez-Ayúcar et al. 2018; Table 1), visited comets using fast flybys of many km/s through a comet's coma at distances of 200 to 20,000 km over the course of a few hours. By contrast, like *Rosetta,* future rendezvous, landing, and sample return missions will be orbiting or station-keeping inside cometary comae with speeds relative to the cometary nucleus on the order of cm/s, i.e., ~$10^5$ times slower than for a flyby (Fig. 4). Such missions will approach to within a few km, or a few nucleus radii, from the comet nucleus' center - and will do this for months at a time. The main hazards for these rendezvous spacecraft are not high-speed impacts of a large (~ 1g) coma dust particles on a vital bus hardware components; instead they are the long-term effects that cometary outgassing can produce near the nucleus' surface on a spacecraft's Attitude Control Systems (ACS), via accumulation of stochastic non-inertial, non-solar forces, potential confusion of spacecraft navigation sensor systems, and potential contamination of spacecraft surfaces exposed for long durations.

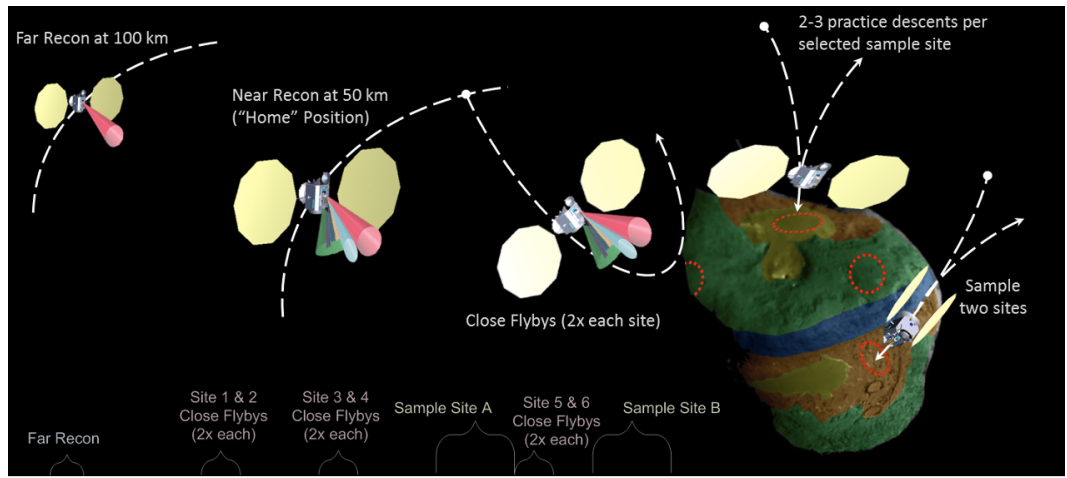

**Figure 4 – Typical stations in the near-nucleus environment anticipated for a state-of-the-art comet rendezvous & "Touch and Go" (TAG) sampling mission spacecraft. A very similar scenario was implemented by the Rosetta mission for Philae site selection and landing procedures. [Image courtesy of APL and Scott Sandford.]**

As will be shown below, detailed modeling, coupled with experience from planning and executing the spacecraft proximity operations of *Rosetta* (Vallat *et al.* 2017, Pérez-Ayúcar *et al.* 2018, Pineau *et al.* 2019) indicate that the near-nucleus environment can be a relatively safe region in which to operate. Gas densities a few meters above a comet's surface are similar to those found in good laboratory vacuums, and the *Rosetta* spacecraft ACS accounted for the stochastic forces with only tens of meters' placement error (Accomazzo *et al.* 2016, 2017). Similarly, surface contamination issues were minimal for *Rosetta* even after 2 years of operations that included excursions to within a few km of the surface. The only significant issue encountered by *Rosetta* was confusion of the celestial star trackers by sunlit dust particles streaking, tumbling, and flying past the spacecraft (Accomazzo *et al.* 2016, Grun *et al.* 2016; see Section 5.1) — a problem that can be easily addressed in future missions using the next generation of star trackers and implementing improved transient rejection algorithms (Accomazzo *et al.* 2017).

## 2.      Physics and Chemistry of the Near-Nucleus Environment.

Spacecraft operating in the dynamic near-nucleus region of a comet experience forces due to cometary gravity, solar radiation pressure, outflow of cometary dust and gas, and drag forces due to spacecraft motion through the coma. It is important to consider the magnitude and direction of each of these forces using realistic engineering models describing the asymmetric outflows (due to sunward emission of material, concentrated jet outflows, and potential extended emission, where ice grains in the coma emit significant amounts of water) as well as the change in total outflow rate as the comet orbits the Sun.

Qualitatively, a near-nucleus spacecraft that is station keeping or slowly maneuvering (< 10 m/s) with respect to the comet nucleus will be operating under the major forces of cometary gravity and solar radiation pressure. In the far field (typically defined as distance $d > 20$ $R_{nucleus}$), these forces are slowly varying on orbital timescales of weeks to months; in the near nucleus region, the irregular shape of the rotating nucleus (known rotation period $P_{rot} = 4$ hrs to 4 days range) can require additional trajectory calculation, trimming and position adjustment on hourly to daily timescales. Typical locations for spacecraft are in 50 or 100 km radius "mapping" orbits oriented to straddle the nucleus's terminator and observe the body's surface as it rotates under the spacecraft, in highly elongated elliptical orbits with few km pericomet center distances, or in neutral "station-keeping" hovers over single nucleus locations like the sub-solar point. In addition, the spacecraft is moving through a very sparsely filled vacuum "cloud" of cometary material (mbar to nbar) streaming away from the nucleus; this stream gently pushes it away from the nucleus while also buffeting it slightly and bringing the occasional dust particle to the spacecraft (Sections 3 and 5).

To get a sense of the most extreme gaseous environment that might be encountered around a comet, we perform the following simple calculation. The gas mass density in the coma near the surface of a 1 km radius, very active comet emitting $Q_{gas}=10^{29}$ molecule/s (=3000 kg/s) of water moving at $v_{gas}$ = 0.5 km/s from its sunlit hemisphere is given by

$$\rho_{gas} \;=\; Q_{gas\frac{(m_{gas} \times N_{avg})}{2\pi r^2 v_{gas}}}$$

$= 10^{29}$molecule/s (18g/mole)/($6.0 \times 10^{23}$ molecules/mole)] /[$2\pi 10^{10} cm^2$ (0.5 x $10^5$ cm/s)]

$= 1$ x $10^{-9}$g/$cm^3$

producing mass fluxes $\rho_{gas}v_{gas}$ of 5 x $10^{-5}$ g/cm$^2$/sec. For comparison, on Earth, the air density at Sea Level is ~ $1.3 \times 10^{-3}$ g/cm$^3$ (~$10^6$ times higher than our extreme coma value), and a light breeze moves at 0.005 km/sec producing effective mass fluxes of 0.65 to 6.5 g/cm$^2$/s. Thus, the gas outflow at the surface of a strong comet is 13,000 times less than that of a light breeze on Earth. If the comet gas were at the ~300 K of Earth sea level, the ideal gas law P = (N/V) RT tells us that this density would be equivalent to a pressure of 8 x $10^{-2}$ Pa, similar to the vacuums created in laboratory research chambers by roughing pumps. In comets, the actual gas temperatures can vary from 30 K to 300 K, so this is an upper limit for the local pressure. Thus, comet outgassing wind buffeting effects will be small, especially compared to the structural design strength of modern spacecraft. Gas contamination effects will be similar to those found in moderate pressure vacuum chambers – surface monolayers of chemi- and physi-sorbed gas (mostly water, carbon monoxide, and carbon dioxide) – but little else.

As with the gas outflow, we can quickly explore the limits of dust particle density at a comet nucleus' surface. For the highly active comet described above, we assume a typical Dust/Gas mass ratio = 3 (Sykes *et al.* 2004, Choukroun *et al.* 2020), in which the comet will be releasing $dM_{dust}$/dt ~10,000 kg/s of dust.  In a worst-case scenario, most likely to apply near a comet's perihelion, we assume that this entire mass takes the form of a small particle dominated cometary dust size distributions with dn/da ~ $a^{-3.8}$ (where a = emitted dust particle radius, running from 0.1 μm to 1 cm; typical comets range from -3.0 in the exponent for sparse large particle dominated dust emission to -3.8 for copious small particle dominated emission; Lisse *et al.* 1998). For dust accelerated via gas drag, $v_{dust}$ ~ $v_{gas}$ sqrt($Q_{gas}$ /$10^{28}$ mol/sec x 0.1 μm/$a_{dust}$), where $a_{dust}$ is the dust particle radius (Ip & Mendis 1974; Delsemme 1982; Krasnopolsky 1987; Vaisberg *et al.* 1987; Lisse *et al.* 1998; Thomas 2020, eqn. 4.101 and references therein), and $m_{dust} = 4/3\pi\rho a_{dust}^3$ , the maximal $v_{dust} = v_{gas} = 0.5$ km/s is found for the smallest, least massive $\sim 0.1$ particles (the smallest nm-sized dust particles act essentially as very large gas molecules), but the maximal momentum transfer goes as $a_{dust}^{2.5}$. Given that the number of particles falls off as $dn/da \sim a^{-3.8}$, we have $d(mv)/da \sim a^{-1.3}$, and the momentum transfer to spacecraft is dominated by impacts from the smallest particles (as assumed in the previous paragraph). Large $(10 - 100$ µm) particles may carry $10^3$ to $10^6$ times more mass than the abundant $\sim 1$ µm particles, but there just aren't very many of them ($10^4$ to $10^8$ less) and they are moving very slowly at 10's of m/s.

We assume porous fluffy aggregated rock-organic dust particles with an average density law of $\rho = 2.5$ g/cm$^3$ (0.1 µm/a)$^{0.1}$ (Lisse *et al.* 1998) and mass $4/3\pi\rho(a)a^3$, and write

$$dM_{dust}/dt = \int_0^\infty C \, (dn/da) \, m(a) \, da = 10^7 \text{ g/s}$$

and

$$dN_{dust}/dt = \int_0^\infty C \, (dn/da) \, da$$

where $dN_{dust}/dt$ is the total number of dust particles emitted per second and C is a normalization constant to be solved for. Resolving the integrals, we find

$$\frac{dM_{dust}}{dt} = (4\pi C/3) \, 2.5 \; g/cm^3 \, (10^{-5} cm)^{0.1}[(1)^{0.1} - (10^{-5})^{0.1}]$$

$$= \left(\frac{2.3g}{cm^{2.8}}\right) C \; = \; 10^7 g/sec$$

Therefore

C $\qquad = 4.4$ x $10^6$ cm$^{2.8}$/s.

This implies that the total dust production rate for the comet would be

$$\frac{dN_{dust}}{dt} \; = \; (4.4 \; x \; 10^6 cm^{2.8}/s) \; /2.8 \; [(10^{-5} cm)^{-2.8} - (1cm)^{-2.8}]$$

$$= 1.5 \; x \; 10^{20} particles/s$$

with the vast majority of the particles being submicron in size.

In order to compare these dust parameters to terrestrial equivalents, we need to compare to particles that are larger than typical coma particles, i.e., particles with a > 0.5 μm. We calculate the number of particles emitted with a > 0.5 μm for our highly active, small particle emitting, high number density comet, and find 1.6 x $10^{18}$ per second. Assuming our comet has a 1 km radius nucleus as before, and dust with surface emission speeds up to 0.1 km/s, these will have 0.5 μm particle space density ~ $(dN_{dust/dt})/4\pi r^2 v_{dust}$ = 1.7 x $10^{18}$ particles/sec /[$4\pi(10^5 cm)^2(1x10^4$ cm/s)] = 1320 particles/$cm^3$ at the nucleus surface. This should be compared to the typical > 0.5 μm particulate count for outside urban air of ~35 particles/$cm^3$ and for room air at ~8 particles/$cm^3$. As the particles expand outward into the nucleus, densities drop off as $1/r^2$, rapidly reducing the amount of dust encountered by a spacecraft farther from the surface – e.g., at just 10 km from the nucleus, the dust particle density in our example will be reduced 100-fold, down to levels of just 13 particles per $cm^3$.

The encountered dust flux can be further reduced by backing away from the comet when it is most active (typically nearest perihelion), and only working in the near-nucleus environment when dust emission is dominated by slowly moving (few m/s), large (0.1 mm - dm) chunks and flakes of nuclear material. This was the strategy implemented, for example, by Rosetta during its 2 years long rendezvous with 67P/Churyumov-Gerasimenko (Pineau *et al.* 2019). In this case we can still follow the same analysis steps as above, but instead assume a more "normal" quiescent level comet emission behavior with dn/da ~ $a^{-3.0}$ (exponent ranges of 2.2 to 3.1 were seen by Rosetta, Fulle *et al.* 2016), so that dust effects become even more benign. In the -3.0 "worst" heavy particle case the total 0.1 μm to 1 cm emitted particle number drops by 2 orders of magnitude to 6 x $10^{18}$/s, with only 7 x $10^{16}$ particles > 0.5 μm emitted per second, ~0.2 particles/$cm^3$ at the nucleus surface, and just 0.002 particles/$cm^3$ at 10 km distance, better than Class 10,000 clean room levels.

However, comets outgas in a heterogeneous and non-uniform manner. The majority of the nucleus surface produces low levels of roughly steady emission, but these are punctuated by very localized regions of enhanced emission (aka "jets") that that can vary on timescales of minutes and be orders of magnitude more active per unit surface area than the median surface activity. For example, on the nucleus of comet 67P/Churyumov-Gerasimenko, the target of the ESA Rosetta long duration rendezvous and landing mission, dust activity was often found to depend on the general location of sources (Schmidt *et al.* 2017). In the northern "winter" hemi-nucleus dust sources tended to be many widely dispersed single dust jets. In contrast, in the equatorial and southern hemi-nucleus regions the dust source was much more broadly distributed – and it was the southern hemi-nucleus which was in summer throughout the perihelion period and dominated the orbital averaged dust emission activity.

Allowing for the 30-40% of emission that happens in some comets from a few surface jet sources (Sekanina *et al.* 2004, Farnham *et al.* 2013), we find that even for a strong dust jet on the surface which concentrates 10% of the comet's dust emission (1000 kg/s) into a narrow region of ~100m radius (0.0025 of our hypothetical 1 km radius comet's surface area) the largest dust jet density we can expect would be about 40 times larger than the bulk average emitted density, or ~2560/cm$^3$ > 0.5 μm at the surface and 26/cm$^3$ at 10 km out (assuming an dn/da ~ a$^{-3.5}$ PSD). Given that good clean rooms for s/c fabrication are classified as Class 100 (defined by US FED STD 209E as < 100 particles of 5 μm radius or greater per ft$^3$ [= < 0.0035 particles/cm$^3$]) to Class 1000 (defined by US FED STD 209E as < 1000 particles of 5 μm radius or greater per ft$^3$ [= < 0.035 particles/cm$^3$]), we can see that the dust density in the near nucleus environment is unlikely to be of major concern for short mission times in these regions. Indeed, the dust density of > 5 μm particles in these jets at 10 km out is lower than that found in a Class 1000 cleanroom!

It is important to emphasize that the previous discussion is for an extreme worst case. Most comets targeted by missions such as a comet surface sample return will be less active by factors of 10 to 100 during the near comet operation phase.

## 3.     Previous Engineering and Analysis Studies.

### 3.1     Byram *et al.* (2007) Engineering Study. One of the key theoretical engineering studies published in the pre-*Rosetta* comet rendezvous era was that of Byram *et al.* (2007), "Models for the Comet Dynamical Environment". Carefully laid out in great detail and still highly relevant post-*Rosetta*[1], we spend some time here reviewing its important highlights, as they

---

[1] The Byram *et al.* 2007 study has been updated recently, by Moretto & McMahon 2020a, who showed that they could predict the orbit evolution due to outgassing even with non-spherical/asymmetric coma models, and by Moretto &

provide critical insight into the important physical operative mechanisms at play in the near-nucleus environment. E.g., the abstract for this paper is indicative of how this work has set the scene for this and similar works: *"An outgassing jet model is presented in support of spacecraft navigation for future missions to comets. The outgassing jet is modeled as an emission cone while the comet is modeled as a uniform density triaxial ellipsoid. The comet's motion about the sun is included in the model. The model is used to explore the effects on a spacecraft passing through an outgassing jet field. The outgassing jet model is also used for simulation and estimation of the physical outgassing properties of jets at and near the surface of a comet. Methods for estimating the locations and sizes of multiple outgassing jets are presented."*

Most useful for this work is Byram *et al.*'s discussion of the expected forces on a spacecraft within a few nucleus radii of an active comet (81P/Wild 2 in the case of Byram *et al.*; the outgassing models are tailored as closely as possible to the 20-some jets estimated to be emanating from the comet by Sekanina *et al.* 2004, and for the fine details of the outgassing rate with heliocentric distance of Wild 2). Wild 2 has a similar activity level to comets typically selected for comet rendezvous missions ($\sim 10^{27} - 10^{28}$ mol/s in the range $1 < r_h < 3$ AU), so the results provide a valid study for general use. Byram *et al.* use the formulation

$$a_p = Q_j \, V_g / B \; (R_{nuc}/r_{jet})^2$$

to estimate the outgassing accelerations on a near-nucleus spacecraft, where $a_p$ = the spacecraft's acceleration, $Q_j$ is the total mass flux of material emitted by the comet, $V_g$ is the gas outflow speed, B = mass/surface area ratio for the near-nucleus spacecraft, $R_{nuc}$ is the radius of the comet, and $r_{jet}$ is the distance of the spacecraft from the radial center of the jet outflow (which may or may not be coincident with the center of mass (COM) of the nucleus). Reproduced here in the black solid curve of Fig. 5 is what Byram *et al.* 2007 predicted as the maximum acceleration experienced by a spacecraft at 1 $R_{nuc}$ (i.e., the limb) from a comet releasing $Q_j = 944$ kg/s :

McMahon 2020b and Scheeres & Marzari (2020) showing that a spacecraft can successfully estimate gravity and gas forces and navigate using on-board sensing & attitude control (OpNav) systems in such arbitrary gas distributions.

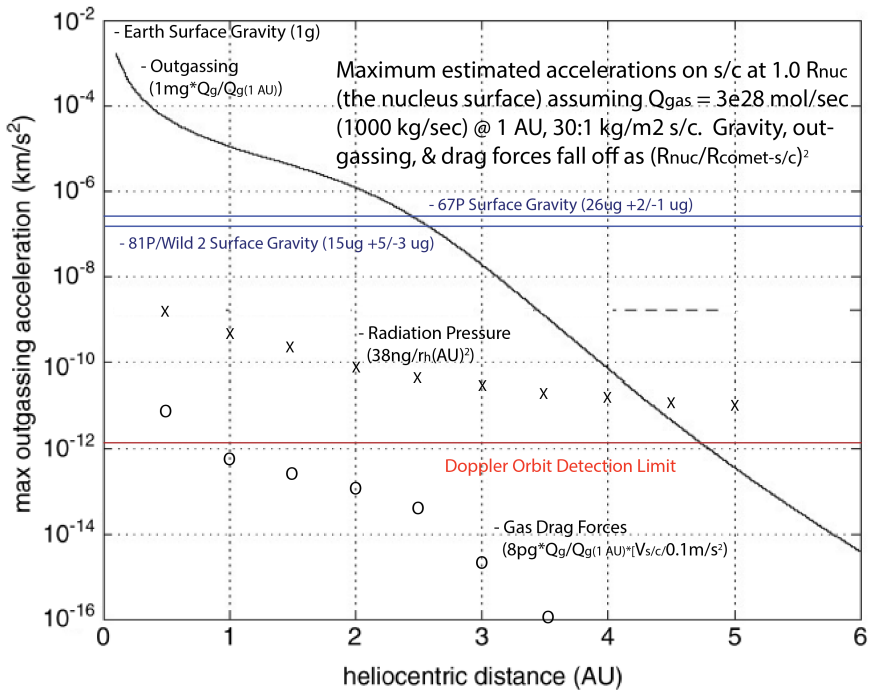

**Figure 5 - Maximum estimated accelerations vs. heliocentric distance on a 40 m² (mainly in solar panels), 1200 kg spacecraft caused by outgassing and gas drag forces, at 1.0 R_nuc.** Original model of Byram *et al.* 2007, Fig. 9 tuned to match the *in situ* measurements of the STARDUST spacecraft at Comet 81P/Wild2, with R_nuc = 2.1 km, in 2004. The total mass outflow rate of dust and gas assumed was 944 kg/s at Wild2's perihelion distance of ~ 1.5 AU. Worst-case solar radiation pressure and gas drag forces are estimated assuming a flat plate s/c with perfect reflectivity and albedo = 0.1. Also shown for comparison are the estimated mean gravitational accelerations at the surfaces of comets 81P and 67P.

In a follow-on study conducted in 2010 – 2012, our group added some important additional estimates to Byram *et al.* 2007's Figure 5 that are shown in our Fig. 5 above. Firstly, we added the expected sensitivity for modern spacecraft orbital Doppler acceleration determination: ~ $10^{-12}$ km/s² ($10^{-10}$ g or 0.1 ng), in the range of the outgassing forces expected during close proximity operations. Secondly, we added the expected force due to solar radiation pressure on a spacecraft. Using $P_{rad}$ = 5.0 μPa (μN/m²) /$r_h$(AU)² for an average albedo = 0.1, solar panel dominated s/c, and assuming a 1000 kg spacecraft with a surface area of 40 m², we have

$$a_{rad} = 5.0 \times 10^{-6} \text{ N/m}^2 \times 40 \text{ m}^2 / r_h(AU)^2 /1000 \text{ kg} = 2.0 \times 10^{-7}/r_h(AU)^2 \text{ m/s}^2$$

Note that the solar radiation forces do not vary with the comet's gas or dust mass production rate, and can dominate the forces on the spacecraft for large spacecraft-comet distances or for very low $Q_g$ rates when the comet is far from the Sun.

Finally, we added to Fig. 5 an estimate of the drag forces for a co-orbiting spacecraft moving at 0.1 m/s (10 cm/s) through a gas coma. Using the equation developed earlier for the gas mass density, $\rho_{gas}$, above a comet nucleus' surface, we have for a total sunlit hemisphere gas production rate of $Q_{gas} = 3x10^{28}$ molecules/s = 1000 kg/s outgassing rate at outflow velocity of 0.3 km/s from a nucleus surface of radius of 2.1 km, a gas mass density of

$$\rho_{gas} = Q_{gas} / (2\pi v_{gas} r_{Nuc}^2) = (10^6 \text{ g/s})/(6.3 \times 0.3 \times 10^5 \text{ cm/s} \times 4.4 \times 10^{10} \text{ cm}^2)$$
$$= 1.2 \times 10^{-10} \text{ g/cm}^3 \text{ (or } 1.2 \times 10^{-7} \text{ kg/m}^3)$$

right above the surface. Assuming unit drag coefficient, the drag forces and accelerations on a 40 $m^2$, 1000 kg s/c moving right above the limb at relative velocity 0.1 m/s are then

$$F_{drag} = \rho_g v_{s/c}^2 A_{s/c} = 1.2 \times 10^{-7} \text{ kg/m}^3 \times (0.1\text{m/s})^2 \times 40 \text{ m}^2 = 4.8 \times 10^{-8} \text{ kg m/s}^2$$

and

$$a_{drag} = 4.8 \times 10^{-8} \text{ kg m/s}^2 / 1000 \text{ kg} = 4.8 \times 10^{-11} \text{ m/s}^2 \text{ for a 1000 kg spacecraft.}$$

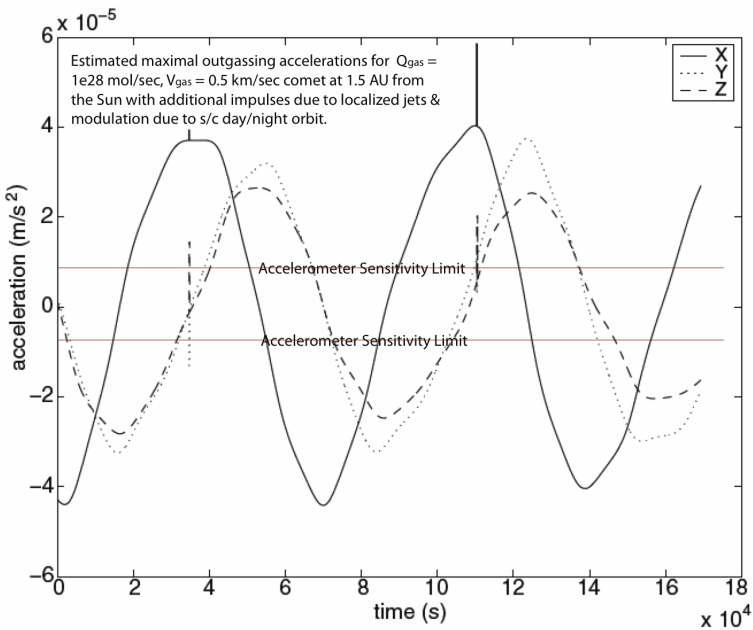

**Figure 6 - Estimated outgassing accelerations on a model 40 m², 1000 kg spacecraft at 1.0 $R_{nuc}$ at 1.5 AU from the Sun, assuming $Q_{gas} = 1x10^{28}$ mol/s and $V_g = 0.5$ km/s outflow velocity.** (note change in vertical scale to $10^{-5}$ m/s² from previous figure). The large scale modulation is due to spacecraft orbital passage over actively outgassing dayside regions and quiescent nightside areas. Orbital period is ~20 hrs. Sensible perturbations due to jet crossings can also be seen at ~4 x 10⁴ and 11 x 10⁴ s. (After Fig. 10 of Byram *et al.* 2007.)

The magnitude of these drag forces is small enough that they can be neglected for short-term operations. Although the accumulations of these forces will affect the long-term motions of the spacecraft, their effects can be monitored and accounted for as needed using an ACS system capable of responding to and making changes on seconds to minutes timescales (Fig. 6).

Given the magnitude of all these effects from their engineering model , Byram *et al.* (2007) made the following mission ops recommendations for dealing with the dust and gas hazards and expected forces on spacecraft in the near-nucleus environment:

- Repetitive approach (>500 km) imaging to obtain rough location of any major jets

- Repetitive approach (>500 km) imaging to obtain timing of any regular flares or outbursts

- Gas sensors to detect important increases in total gas density and comet output ($Q_{gas}$)

- At the closest approach distances, a spacecraft ACS with 0.5 ng sensitivity able to detect the accelerations due to outgassing at r < 3 AU, and due to major jets at r < 2 AU

- Cometary outgassing dominates cometary gravity for r < 2.5 AU (assuming a water sublimation dominated, > 1 km radius nucleus), requiring ***active*** guidance at closest approach

- Predictable and controllable radiation pressure forces dominate outside 4 AU, and can dominate for spacecraft-comet distances > 30 $R_{nuc}$ (~15 km) at 3 AU, > 100 $R_{nuc}$ (~50 km) at 2 AU, and 300 $R_{nuc}$ (150 km) at 1 AU

## 3.2 Rosetta Pre-Encounter Models.

The *Rosetta* rendezvous and lander mission to comet 67P/Churyumov-Gerasimenko faced similar issues. The *Rosetta* dust modeling team of E. Grün *et al.* spent some time studying and modeling the expected environmental conditions around 67P/C-G, and came to similar conclusions (Agarwal *et al.* 2007) as the coarse models given above in Section 2 – that the main hazard to *Rosetta* was due to asymmetrical forces, as well as the possible buildup of dust coatings on the spacecraft, and that mitigation provided by instrument covers and baffles was necessary. The exact details of the dust hazard in their models, though, was found to be highly dust particle size distribution (PSD) dependent, especially to the slope of the distribution (typically between -3.0 and -4.0), which controls the relative number of small vs. large particles, and thus whether the outflowing dust's mass is concentrated in the smallest micron sized particles (dn/da ~ $a^{-4}$) or in the largest cm to dm sized particles. Recall from Section 2 that the speed of ejection of dust particles from an active nucleus goes roughly as the inverse ½ power of their size, and that in the micron sized limit, they act like the gas, flowing out at v ~ $v_{gas}$ (Finson & Probstein 1968, Lisse *et al.* 1998, 2004, Rinaldi *et al.* 2018), while for the largest emitted particles on the order of 1 cm or so, they move away from the nucleus at about the escape velocity, or ~1 m/s (as confirmed by in situ s/c; Kelley *et al.* 2013, Agarwal *et al.* 2016). Thus while the largest particles that dominate low level, steady dust comet emission (Lisse *et al.* 1998, 2004, Fulle *et al.* 2016) can have masses billions to trillions times the smallest, they move 100's of times slower and are billions to trillions of times rarer, making a dn/da $\sim$ a$^{-3}$ "rarified snowball" dust coma much less difficult for a spacecraft to navigate through than a dn/da $\sim$ a$^{-4}$ "tiny particle sandstorm" coma. (Rosetta actually encountered both kinds of dust emission , and remained far from 67P's nucleus when the comet was within $r_h = 2.5$ au of the Sun and highly active; See Section 4 and (Fulle *et al.* 2016, 2018; Levasseur-Regourd *et al.* 2018; Güttler *et al.* 2019; Longobardo *et al.* 2020) for more details of this experience and Section 5 for more details on expected dust-s/c effects and interactions.)

## 4.    Spacecraft Experience in the Near-Nucleus Environment: Results From the (*Giotto, Vega1/2, DS-1, Deep Impact, Stardust, & Rosetta*) Missions

## 4.1    The Giotto, Vega1/2, DS-1, Deep Impact, and Stardust Fast Flybys.

Although the early spacecraft explorations of comets were all done as fast flyby reconnaissance missions that approached to within only 200 to 8,000 km of their respective targets, we can evaluate the risks involved in those events and compare them to the concerns relating to near-nucleus environment rendezvous and landing missions. The Giotto mission flew by two comets, a 68.4 km/s flyby within ~600 km of prime target 1P/Halley in 1986, and a 14 km/s flyby 200 km from secondary comet 26P/Grigg-Skjellerup in 1992 (Table 1). Vega 1 & Vega 2 flew by 1P/Halley in 1986, both at distances ~8,000 km. Deep Space 1 flew within 2200 km of comet 19P/Borrelly in 2001. In 2005, the impactor portion of the Deep Impact spacecraft reached the nucleus of comet 9P/Tempel 1, while the flyby portion viewed it from a distance of 500 km. The flyby portion then continued on to fly within 700 km of comet 103P/Hartley 2 in 2010. The Stardust spacecraft encountered comet 81P/Wild 2 at a distance of 240 km and relative speed 6.2 km/s in 2004, and then visited comet 9P/Tempel 1 at 180 km closest approach in 2011, one orbit after the Deep Impact excavation experiment.  (Two other spacecraft, Suisei and Sakigake also explored comet Halley, but their encounter distances were much larger than those listed above, and thus the risk of cometary debris was negligible.)

These missions carried a variety of instrument suites, but due to the high speeds of the encounters, all were concerned with the effects of hazardous dust impacts and all but DS-1 carried some form of impact shielding. Of the 8 fast flybys, 3 encountered effects due to large dust particle hits: the Giotto s/c experienced a large impact impulse near its closest approach to the Halley nucleus of ~600 km that destabilized its spin (Curdt & Keller 1988, Jessberger & Kissel 1991, McDonnell *et al.* 1991), and likely also destroyed its multicolour imaging camera functionality; the Deep Impact impactor section experienced at least two large impulses (as recorded in camera images) within 100 km of 9P's surface; and the Deep Impact flyby s/c endured ~20 ACS reset events as it flew within 700 km of hyperactive comet Hartley 2. None of these impacts affected the survivability of the respective s/c, and only the first-ever comet flyby mission, ESA's *Giotto*, whose dust impact energy was more than 30 times (~$v^2$) and dust impact probability was ~50 times greater that of the other fast flyby missions because of the huge production rate from 1P/Halley, suffered significant hardware damage.

It is important to note that, because of the high encounter velocities for these missions, the dust outflow velocities from the comet nucleus were close to negligible; the relative velocities of the spacecraft and comet nucleus set the scale for any momentum transferred by impacts. Similarly, neutral gas effects were not important at the flyby distances (given the short duration of the pass), and the plasma environment found around comets in the early flybys was found to be more rarified and benign than the LEO environment regularly worked in by the Space Shuttle, Skylab, MIR, ISS, etc.

## 4.2 The *Rosetta* Rendezvous and Landing Mission.

The *Rosetta* mission at comet 67P/Churyumov-Gerasimenko (hereafter 67P) represents the first ever long-term rendezvous and landing mission to a comet (Figs. 1, 3, and 7). As such, it allowed us to take previous theoretical and model predictions and update them for actual *in situ* measurements in a near-nucleus environment for a comet that emitted as much as 2 x $10^{28}$ mol/s of gas and its entrained nm to dm sized dust particles (Fig. 8)[2].

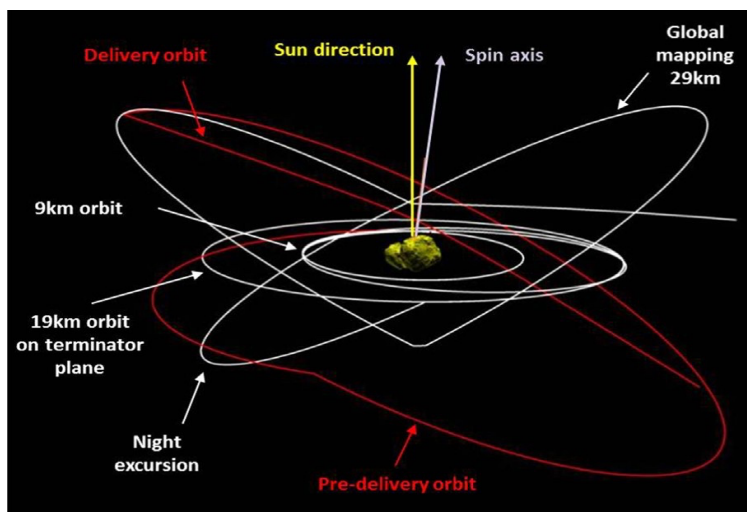

**Fig. 7 - The Rosetta mission's encounter, capture and global mapping orbits for comet 67P. The red curves labeled "Pre-delivery orbit and "Delivery orbit" are the end of the launch and planetary flyby solar system trajectory shown in gross scale in Fig. 1** (After Accomazzo *et al.* 2016.)

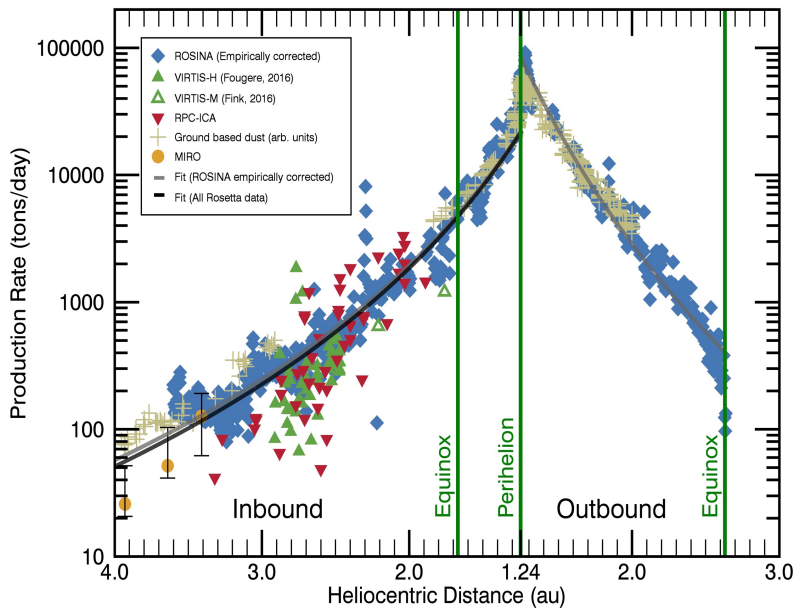

**Fig. 8** – $Q_{gas}$ vs $r_h$ for 67P as measured by multiple Rosetta instruments (Hansen *et al.* 2016). Rosetta's target comet 67P increased its outgassing emission rate more than 3 orders of magnitude during the course of the rendezvous mission that took the spacecraft + comet through perihelion. (1 ton/day ~ 2.9 x $10^{23}$ mol/s assuming <M.W.emitted> = 24 amu.)

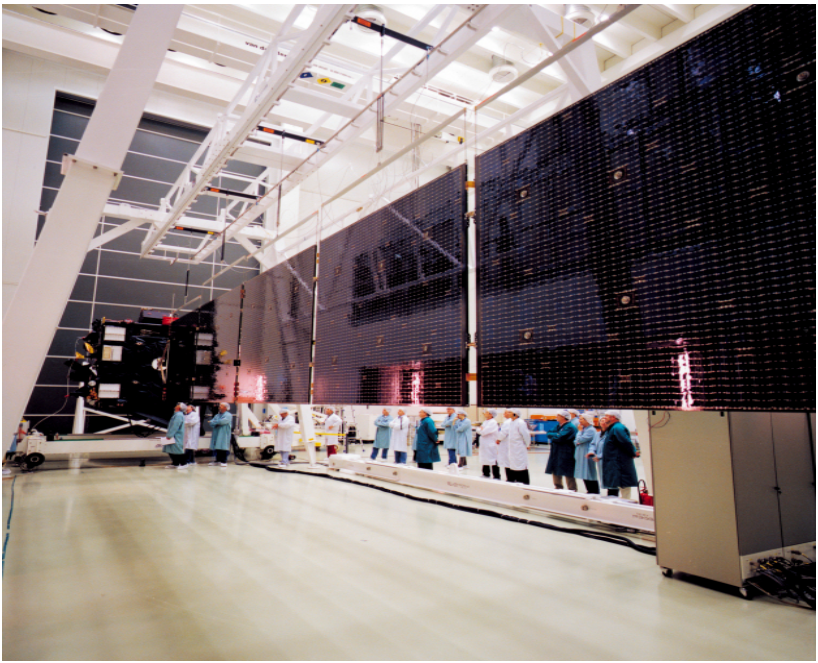

**Figure 9 - Assembly clean room image showing the spacecraft's cubic main bus and the large scale of one of Rosetta's 14 m long solar panels**, required in order to develop sufficient on-board power at > 4 AU from the Sun. (Image from ESA and A. Van der Geest, https://sci.esa.int/web/rosetta/-/54421-rosetta-solar-panels).

Consisting of a cube of dimensions 2.8 x 2.1 x 2.0 m attached to two fixed, 14-m long solar panels, each of 32 m² surface area extending to either side (Fig. 9), the *Rosetta* spacecraft was decidedly not "aerodynamic" nor was it designed to minimize any effects of gas drag, and thus should have encountered close to the maximal extent of these effects during its time in close proximity to 67P.

---

[2] N.B. - For the purposes of brevity and to maintain the focus of this paper on the operation of s/c around cometary nucleiwe have not gone into the details of the findings of the ROSETTA mission concerning the nature of the dust in the coma of comet 67P. We refer the reader to the following excellent detailed mission results reviews and the references they contain for more dusty information: Fulle *et al.* 2016, 2018; Güttler *et al.* 2019; Levasseur-Regourd *et al.* 2018; and Longobardo *et al.* 2020.

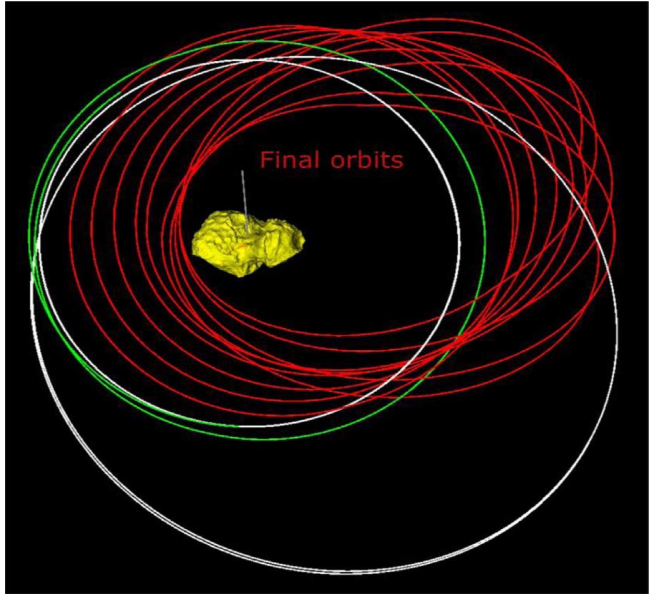

**Figure 10 - Rosetta spacecraft post-perihelion near-nucleus orbits,** including the ones leading to the final touchdown and landing of the main spacecraft on the surface of the comet. (After Accomazzo *et al.* 2017.)

Despite this, the *Rosetta* bus was routinely able to perform precision navigation within 100 km of the 67P nucleus, and made a number of forays and maneuvers inside of 10 km (Fig. 10), while continually measuring moderate acceleration levels on the spacecraft (Figs. 11 & 12) consistent with the predictions of Byram *et al.* 2007 given above. This is a direct consequence of the gas drag effects being small and manageable in the rarified near-nucleus comet environment (Section 1).

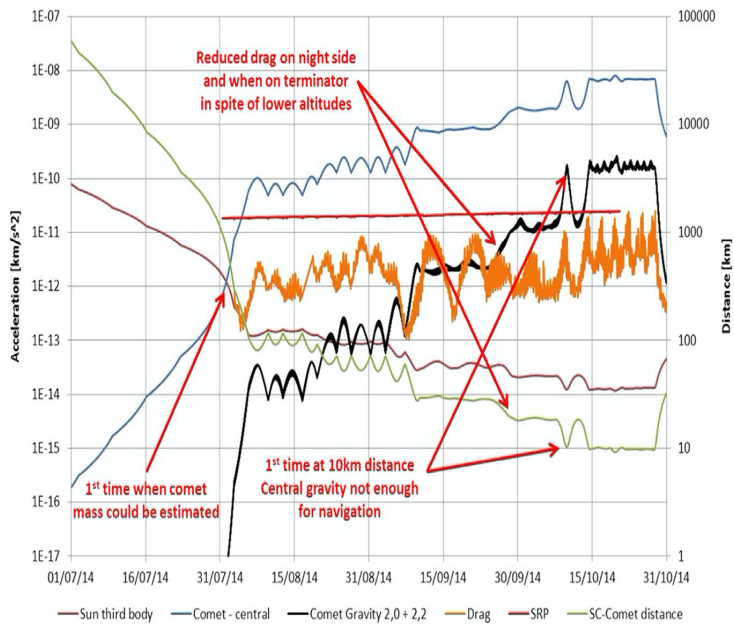

**Figure 11 – Rosetta measured accelerations and spacecraft-comet distance for the initial rendezvous and capture phase of the mission** (July to Oct 2014; after Accomazzo *et al.* 2016). The measured gas production rate for 67P at first spacecraft encounter was ~ $1.7 \times 10^{24}$ molecules/s at $r_h$ ~4 AU [Fig 11]. Compare the magnitude of the red "SRP" spacecraft engine forces required to maintain station keeping and the orange coma "Drag" curve magnitudes to Fig. 9 of Byram *et al.*, now assuming a comet with 1.0 $R_{nuc}$ = 2.0 km, so that for the same outgassing rate the force on a body should be about that for 1.0 $R_{nuc}$ = 2.1 km 81P/Wild2. Total Rosetta s/c surface area, including the solar panels, was ~70 $m^2$, or ~50% larger than the 40 $m^2$ assumed in the model of Byram *et al.* (Fig. 9). Byram *et al.* predicted ~ $10^{-10}$ km/s² accelerations at the nucleus surface, while Rosetta experienced total accelerations ~ few x $10^{-11}$ at 10 km distance from the comet center.

*Rosetta* did experience some navigational problems, but they were caused by the inability of the star trackers to filter out the signal from dust particles near the spacecraft, ***not forces associated with the flow of gas or dust.*** On occasion, the trackers interpreted moving dust particles as drifting stars, resulting in erroneous ACS "corrections" (Buemi *et al.* 1999; Grün *et al.* 2016). Minor star loss would result in navigational position and pointing errors (Fig. 13).

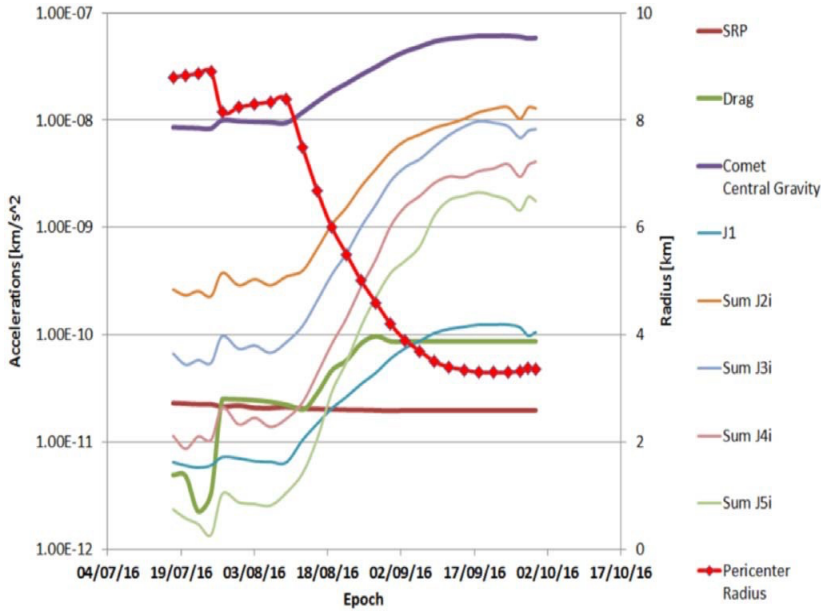

**Figure 12 – Rosetta measured accelerations and spacecraft-comet distance for the post-perihelion and final main spacecraft landing phases of the mission** (April to Oct 2016; after Accomazzo *et al.* 2017). Note that the final landing at ~3.2 km pericenter distance occurred at $r_h \sim 4.2$ AU. The measured gas production rate for 67P at first s/c encounter was $\sim 1.7 \times 10^{24}$ molecules/s at $r_h$ ~4.2 AU and the total measured gas drag accelerations were found to be $\sim 10^{-10}$ km/s$^2$ at the comet's surface, consistent with the predictions of Byram *et al.* 2007.

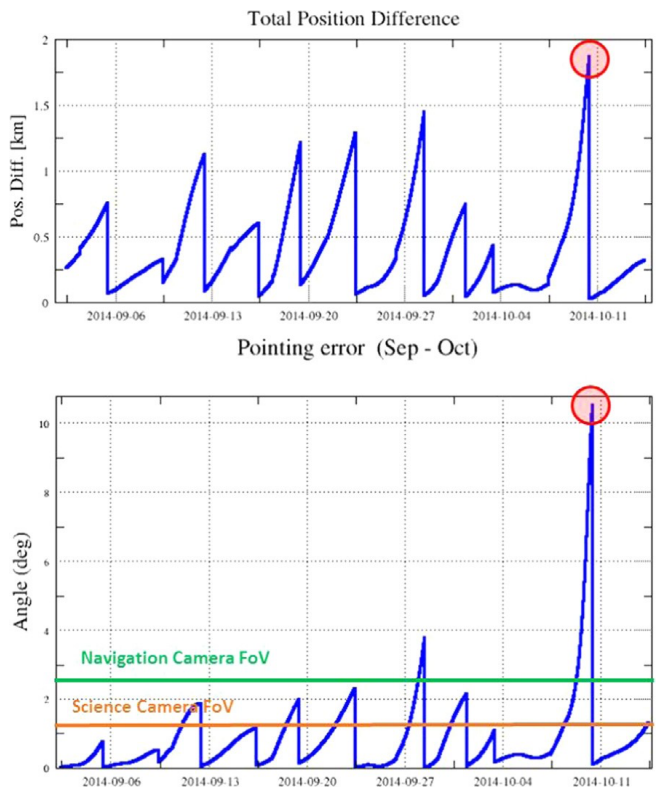

**Fig. 13 - Position and pointing errors encountered by the Rosetta spacecraft** when conducting a maneuver down to 9 km distance in the early post-capture phase of the mission. (After Accomazzo *et al.* 2016.)

However, two major errors resulted in pointing drifts so large that the spacecraft antenna lost pointing lock on the Earth, and the spacecraft went into safe mode. *Rosetta* mission ops dealt with this issue by moving away from the nucleus, with standard operations at 400 km. Because dust is a concern in any near-nucleus mission, such star tracker problems must be alleviated in future missions. Fortunately, the next generation of higher-sensitivity star trackers, combined with intelligent algorithms that can reject moving and transient sources, provides a straightforward solution (see, for example, https://spinoff.nasa.gov/Spinoff2019/it_2.html and https://adcolespace. com/product/adcole-space-star-tracker/ or https://blog.satsearch.co/2019-11-26-satellite-star trackers-the-cutting-edge-celestial-).

It should be noted that the ESA *Rosetta* mission teams did not report any obvious deleterious effects on the operations of the WAC, NAC, and OSIRIS optical cameras and the spacecraft solar panels due to dust buildup on these surfaces throughout the lifetime of the mission; the only mentions of any possible dust effects on spacecraft operations are (1) hypotheses that the "anomalous chameleon" feature seen in ALICE UV spectrometer data (Noonan *et al.* 2016), (2) an unusually high level of noise in one of the GIADA dust analyzer GDS laser systems (Sordinia *et al.* 2019) that could have been due to dust accumulation, and (3) reports of a brief wide spread instrument dust loading at 18h local on September 5, 2016 (day 766 of the mission) when the comet was 3.7 au from the Sun and the s/c was operating within 1.9 km of the nucleus surface; this loading led to the input of a dust grain into the ROSINA mass spectrometer system and the discovery of semi-volatile ammoniated salts in the nucleus of 67P (Altwegg *et al.* 2020).

## 5. Lessons Learned for Future Missions (Close Flyby, Rendezvous, Sampling, Landing); New Analysis/Amelioration for Improving S/C Operations

In this section we discuss the pre-*Rosetta* expectations for comet-spacecraft effects and compare them, when possible, to the direct *Rosetta* mission spacecraft experience in the near-nucleus environment of comet 67P. We also give our best estimates for how the effects should scale with comet activity ($Q_{gas}$), as an aid to the reader for applying the discussion to other target comets. In general, we would expect the effects to scale roughly linearly with comet activity. Thus, in the cases where we reference the *Rosetta* experience with an effect, it should be remembered that this comet reached a maximum observed gas production rate of a little less than $10^{28}$ mol/s at r = 1.24 AU from the Sun.

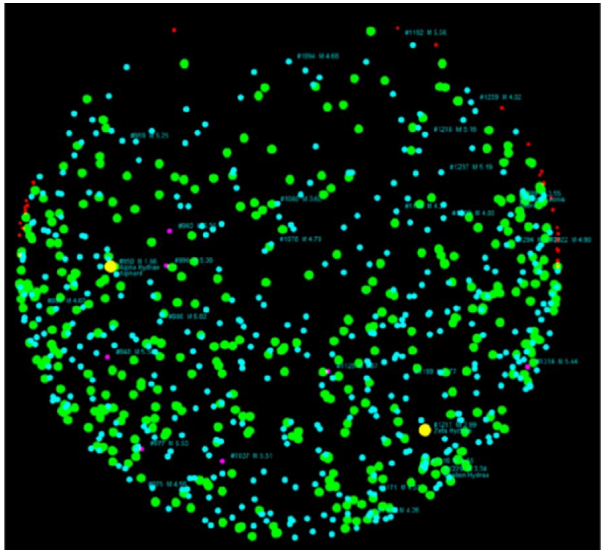

**Fig. 14 - Simulated star tracker field of view.** The fixed stars are shown in red and yellow, the model coma dust grains in green and blue.

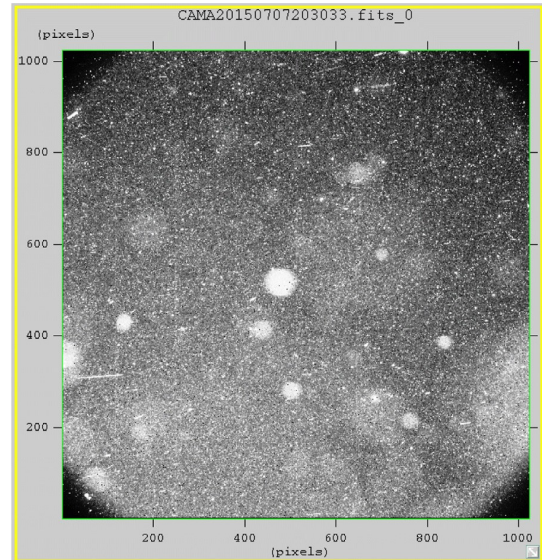

**Fig. 15 – Coma image taken by the navigation camera on 7 July 2015**, showing a sky full of dust.

**5.1 Navigation.** Pre-encounter mission planning studies (Miller *et al.* 1989, Weeks 1995, Byram *et al.* 1997, etc.) focused on the problems of controlled navigation close to a small, rotating, non-axisymmetric body, and in fact *Rosetta*'s biggest problems in operating within the near-nucleus environment of comet 67P from 2014 through 2016 were navigational. The requirement for accurate, updated trajectory information can be seen in Fig. 13. Position errors at the start of nucleus descents grew rapidly and instrumentation would not acquire desired targets unless corrections were made in real-time (~every 10 s, due to the low absolute levels of accelerations, Section 1 and Fig. 3; see Pineau *et al.* and references therein).

However, operational experience during the *Rosetta* mission (Accomazzo *et al.* 2016, 2017) revealed that the prevalence of dust particles in the near-nucleus coma could easily overwhelm *Rosetta*'s active star tracking capabilities (Figs. 14 and 15), as the number of dust particles in the tracker's field of view often exceeded by orders of magnitude the number of available bright navigation stars.

The number of dust particles imaged in an instrument's field of view at any time should scale linearly with the comet's gas production rate $Q_{gas}$, so this effect would be most pronounced for highly active comets at perihelion. Thus due to the star tracker confusion issue, the Rosetta spacecraft was positioned far from the nucleus (at distances > 100 km) during most of the near-perihelion phase, when the heliocentric distance was smaller than 2.5 AU. Dust fluxes resulting from radial outflows decrease as $1/r^2$, hence these large cometo-centric distances much reduced potential surface dust contamination compared to that which would have resulted from near nucleus operations during that period.

To mitigate this issue, instead of navigation with star trackers designed for use in an ACS as the active primary navigation component in order to minimize position and course errors in real time, star trackers should be made secondary navigation tools, used as tie points when the skies are clear far from the nucleus, even with updated modern trackers using the latest moving object rejection algorithms. The density of near nucleus dust will generally be too high versus the available number of bright stars/arcsec$^2$. It is now clear that a better ACS design should be based on the use of gyroscopes with excellent performance for short/mid-term trajectory propagation and the availability of inertial measurement from the star tracker at a frequency that is enough to maintain the drift effects of the gyroscopes within the required limits. Updated star trackers for better false positive detection, utilizing better predictive software algorithms for finding real stars and rejecting non-inertially moving dust particles, quickly and assiduously mapping out the gravity field of the comet environment to 5$^{th}$ or greater order, and optimizing the spacecraft gyroscopes for minimal error drift rates will all be important for the success of future close nucleus spacecraft operations.

**5.2 Mission Planning.** Future missions will also need to allocate significant scientific and engineering manpower resources to mission planning during proximity operations. These resources will need to be applied towards orbit and the attitude planning in a centralized process. Problems due to near-nucleus ACS upsets should be expected, anticipated, and trained for, and safe, robust abort/escape-to-safe/reset navigation procedures developed and implemented in future flight systems. These navigation procedures need to be meshed coherently with the technical goals of the mission and the variable outgassing environment of an active comet (e.g., 67P, Hansen *et al.* 2016, Marshall *et al.* 2019; Fig. 8), so that safe mode implementation rescues the spacecraft with 100% assurance while also minimizing mission impact; e.g., the safe mode's orbital radius needs to be minimized but grow with increasing comet dust emission, and safe orbits need to avoid routinely overflying currently highly active surface regions. Sufficient time, weeks to months, should be allocated to mapping the nucleus's surface in visible and infrared light, as well as in its gas outflow activity, in order to fully understand the surface topography, gravity field, and temporally variable outgassing behavior that can affect near-nucleus spacecraft navigation. Short term (weekly), medium term (bimonthly), and long term (yearly) schedules will need to be maintained in order to cope with science activities keyed to comet behaviors on rotational, seasonal, and orbital timescales. Again the science observations experience gained by Rosetta at comet 67P is an excellent place to start (Vallat *et al.* 2017, Pérez-Ayúcar *et al.* 2018, Pineau *et al.* 2019).

**5.3  Dust Effects: Physical Coverage of S/C; Charged Dust-S/C Interactions; Large Dust Grains.** By contrast with navigational issues, none of these expected problems were found to be of significant concern for the *Rosetta* mission at comet 67P. While the observed high areal density of false positive star tracker objects due to coma dust within a few km of the nucleus is consistent with pre-*Rosetta* model estimates of high dust particle deposition rates onto spacecraft surfaces found in the authors' 2010-2012 follow-up study of Byram *et al.* (2007; Section 3; Figs 14 - 15), the *Rosetta* mission team did not report any issues caused by dust particle coverage, contamination of s/c hardware, charging of the spacecraft, or impacts of large dust grains onto the spacecraft (with the possible exception of potential nanodust effects creating the "Chameleon" anomalous spectral feature of the Rosetta/ALICE instrument, Noonan *et al.* 2016 and the Rosina ammoniated phyllosilicate study engendered by a small particle becoming wedged in the DNFS, Altwegg *et al.* 2020).

The difference in prediction vs. *in situ* experience does, however, highlight the importance of examining the comet prox-ops problem carefully in light of *Rosetta*'s first ever long term rendezvous with a comet nucleus, a mission which included multiple complex trajectories and two landing events on a comet from a year before perihelion to a year after perihelion (Figs. 8 – 11). Rosetta spent, by intentional design, minimal time at $r_{sc-nucleus} < 100$ km of 67P's nucleus once the comet was within 2.5 au of the Sun and the comet's production rates $Q_{gas}$ and $Q_{dust}$ increased by orders of magnitude as bulk water ice began subliming (Fig. 8). Rosetta simply didn't explore this portion of operations phase space. The size distribution of the dust also became much finer and the total emitted dust surface area much larger (Fulle *et al.* 2016, Levasseur-Regourd *et al.* 2018). As the effects of dust coverage, charging, and impacts are all expected to scale as $Q_{dust-surface-area}/r_{sc-nucleus}^2$, it is possible that a future s/c operating near perihelion in the near-nucleus environment for large amounts of time (i.e., days to weeks) will encounter important deleterious effects, although adopting mitigation techniques like implementing instrumental dust covers, continual performance monitoring of optical instruments and solar panels, and bans on instrument boresight pointing along radial lines of sight to the nucleus can help ameliorate these effects.

**5.4    Stochastic & Periodic Outbursts.**    Comets can exhibit occasional quickly increasing outbursts of activity that can multiply their emission output manyfold, and are expected to be independent of a comet's overall average gas production rate $Q_{gas}$. The *Rosetta* spacecraft observed a number of these with duty cycles of a few/year (Grün *et al.* 2016; Hansen *et al.* 2016: Vincent *et al.* 2016; Agarwal *et al.* 2017), but none caused any serious operations issues with the spacecraft. (It would take outbursts multiplying a comet's total mass output by many orders of magnitude to create a worrisome environment with pressures nearing mbar levels; see Sections 1 & 2 and Fink *et al.* 2021's excellent recent engineering treatment of the subject.) Perhaps one comet in the history of modern observations could fit this scenario – 17P/Holmes was seen to brighten by 4 orders of magnitude in late 2007 due to a major fragmentation "calving" event; (Reach *et al.* 2010, Stevenson *et al.* 2010, Stevenson & Jewitt 2013). But this comet was also well known to have exhibited this behavior previously; see F. Whipple (1985) – and so could be ruled out as a possible mission target if this is a concern. This eventuality does highlight, though, the need for adequate remote lightcurve temporal trending of a mission target before its selection. As these transient events generally last only for a few days, in the event that such an outburst is witnessed during a rendezvous the s/c would simply move out to past ~1000 km and just wait until the activity levels die down to sample. Much has been learned about cometary outbursts since Byram *et al.* 2007, for example the episodic landslide model of Steckloff *et al.* 2016 based on *Rosetta* results, but this is another excellent example of a cometary phenomenon that needed to be experienced *in situ* to be understood.

Small periodic outbursts with quick (< 10 min rise times) have also been seen on comets like 9P/Tempel 1 by the Deep Impact s/c– and these short, approximately 2x increases in brightness were attributed to localized icy patches rotating through the dawn terminator into sunlight. As long as a spacecraft was not located within a few meters directly overhead of such a patch when the sun rose over it, these gentler (likely because they regularly blow off excess volatile pressure) type of outbursts would not significantly affect spacecraft trajectories in the near-nucleus region. The effects of stochastic outbursts are expected to be independent of a comet's overall average gas production rate $Q_{gas}$.

**5.5**  **Dust Jet Coverage and Scouring of Spacecraft Surfaces** – The other exception to the benign nature of cometary coma effects onto a nearby operating spacecraft are the highly localized regions of dust jet emission. Poorly understood, these regions can focus upwards of 10% of a comets entire mass outflow flux sourced from a region only about 10 to 100 meters across, and can vary their output by as much as an order of magnitude in less than 10 minutes. Most jets seem diffuse and quickly expand as much tangentially as they do radially in the first few km above the nucleus, but a few have been seen to stay collimated and focused out to 10's of km from the nucleus surface (e.g., Rabinowitz *et al.* 1999, Soderblom *et al.* 2002, Lin *et al.* 2015, Lai *et al.* 2019). Dust speeds in these jets are poorly known, but are likely to be upwards of several hundred m/s at the nucleus surface (i.e., up to $v_{gas} = 0.1 - 0.8$ km/sec for the smallest submicron dust particles, akin to velocities used in terrestrial sand blasting abraders with 10 - 1000 µm grit particles) and are thus to be avoided. Impacts on a spacecraft from dust entrained dust by these jets would mainly be by fine sub-micron to a few micron sized particles which will include icy (wet) and mineral grains.

These impacts would not penetrate spacecraft panels, but they will cover exposed surfaces if the dust sticks. If we use the number density *at the nucleus surface* (Section 2) of 165/cm$^3$ and an in-jet velocity of 300 m/s, or a flux of 4.0 x 10$^7$ cm$^{-2}$ s$^{-1}$, and assume that the average impactor has 0.5 um radius, then we can estimate that for a unit sticking probability that *in 1 second of direct exposure to an oncoming jet* at full velocity that about 40% of the surface is covered to a depth of 0.5 µm in dust (corresponding to $\tau \sim 0.05$, assuming $\tau \sim 1$ for silaceous dust particles of 10 um radius (Lisse *et al.* 1994, 1998, 2004).

Even worse effects are realized if the dust pits the surface it encounters, as each impact has the potential to make a pit up to 10 times the impactor's radius (i.e., *up to* 5 µm for an 0.5 um dust particle). In the *worst* case of 5 µm pit creation, 100% of an exposed surface could become pitted within less than 1 second of exposure, completely clouding it. The dust flux causing these coverage and scouring effects does decrease rapidly with altitude in the jet once it becomes defocused, however, as roughly

$$Flux\ (r_{s/c-nuc}) \sim Flux_{surface}\ x\ (R_{nuc}/\ r_{s/c-nuc})^2$$

so that at 100 km distance from a 1 km radius nucleus it would take a 10,000 s (2.8 hr) stare down a jet's axis to cause the same level of damage to an optical surface as received during a 1 s stare at 1 km distance. (This is why it is safe to reconnoiter the nucleus surface from a 100 km radius "mapping" orbit.)

Given that typical exposure times for optical cameras are on the order of 10 ms – 0.5 s, it is clear that *a spacecraft should not attempt to operate inside a jet region within a few km of the nucleus, nor should it attempt to image or study it with boresight pointed radially down the jet.* Fortunately, strong jet regions are easily seen by the large amount of light scattering dust they entrain, and thus can be located and mapped during the reconnaissance phase of the mission and avoided if necessary during near-nucleus operations - and Rosetta did precisely this, operating free of deleterious jet effects from comet 67P over the entire course of its 2 year near-nucleus operations. If for any reason a future spacecraft inadvertently finds itself passing through a coma jet within a few km of the nucleus surface, any hazards (like clouding of optics or loss of ACS opnav position knowledge) could be ameliorated by including protective covers or shutters for all spacecraft optics and defining a default safe mode whereby the spacecraft acquires and points towards the Sun and removes itself to > 100 km distance.

**5.6    Tangential gas flows (surface winds) and Spacecraft Torques.** Model calculations by Combi *et al.* (1997, 2012) predicted, pre-Rosetta, that gas molecules coming from a localized jet region of activity directed radially (typically more than 10s of meters across) will expand, due to their thermal velocity, laterally into the vacuum above the surface, producing a wind that will act parallel to the surface. This occurs in both limits of high gas density in the fluid case where the pressure expands the jet laterally as well as in the limit of low gas where gas diffuses laterally in free molecular flow. The dust particles in these models near the edge of a gas jet are also pushed tangentially to the side of the jet initially where they are then entrained in a much lower gas density flow where the radial acceleration of the dust is much lower. Therefore, gradients in the gas flux at the surface can produce lateral surface winds from higher flux to lower flux regions that can also push dust particles or spacecraft.

Evidence for tangential gas flow **was** found in the aeolian dune structures on the surface of 67P by Rosetta (Jia *et al.* 2017, El-Maarry *et al.* 2019). However, with the improved understanding from ROSETTA that jets are often associated with landslide features created by surface material failures (and not large fissures, cracks, or pipes channeling gas from the deep interior, as was previously thought), the "opening nozzle direction" of a jet could very easily be perpendicular to the local gravity field and the local surface normal, providing significant initial tangential impetus to the gas outflow - an impetus that will look roughly radial due to simple geometry once the gas has traveled more than a few body radii from the nucleus (e.g., more than 3 km from a 1 km radius nucleus). Estimates for the speed of tangential winds at the comet's surface range from

$$v_{tangential\ jet\ winds} = 0.1\ to\ 0.5\ km/sec\ \cos(\theta_{jet})$$
$$= 1\ x\ 10^3\ to\ 5\ x\ 10^4\ cm/sec$$

and tangential jet forces on a s/c should be significant and on the same scale as those produced by cometary outgassing winds (see Byram *et al.* 2007 estimates, Section 3.1 and Figs. 5 & 6). Relative nuclear rotation versus a radially outflowing coma can also create tangential winds; a rough estimate for 67P with its $P_{rot}$ = 12 hrs and r ~ 2.4 km yields a surface velocity of

$$v_{tangential\ rotational\ winds} = 2\pi\ x\ 2.4\ x\ 10^5\ cm\ (12\ hr * 3600\ s/hr)$$
$$= 35\ cm/s$$

relative to any static gas envelope, a factor of 30 to 1400 smaller than our estimated tangential gas jet velocities, but still ~1/3 the escape velocity from the object and acting over the entire nucleus surface. In this case, noting the consistency of the coma gas drag forces predicted by the Byram *et al.* 2007 model (Section 3) with those measured by the *Rosetta* s/c (Section 4.2), we see from Figs. 5 & 6 that we do not expect the perturbations due to rotationally produced winds on a near-surface s/c to be large, $< 10^{-8}$ g; this is again a direct consequence of the very low ambient surface pressure at the comet.

The average torques on a spacecraft are expected to scale linearly with the comet's gas production rate $Q_{gas}$, and the number of small highly active regions present that could produce transient strong torques to scale more steeply than linear, so this effect would be most significant for highly active comets at perihelion. However, even the smallest jets found on comets are 10s of meters across, so the gradients will not be extreme and modern ACS systems containing reaction wheels and gyros, coupled with fold-up solar panels, will be able to compensate to maintain stability – with some hit to position knowledge certainty.

## 6.     Conclusions

We have presented a study of the current post-*Rosetta* state of knowledge concerning the near-nucleus comet environment and near-nucleus operating spacecraft. Utilizing simple back of the envelope calculations, sophisticated engineering models of spacecraft behavior, and experience from the *Rosetta* mission's long duration operational history at comet 67P, we determine that the near-nucleus environment is a relatively safe region in which to operate, with gas densities similar to those found in good laboratory vacuums and dust densities better than Class 1 cleanrooms.

The strongest effects we expect on future spacecraft in the near-comet nucleus region are produced by outflowing gas emission during the comet's active phase. Designing an ACS gyroscopic system for a modern s/c with 10's of meter$^2$ projected surface area that can maintain proper spacecraft orientation and minimize trajectory errors to within 10's of meters' navigation error, using off-the-shelf hardware, should be straightforward - especially when coupled with intelligent anomaly avoidance trajectories informed by global mapping surveys (i.e., making sure not to fly the spacecraft through a surface region supporting a strong jet; see Rizos *et al.* 2021 for how a very recent asteroid sample return mission used SOTA AI to intelligently map its target body's surface). I.e., the best mitigation technique against loss of position knowledge during close flybys will be for the spacecraft to avoid small localized patches of high outgassing activity found during the global mapping phase, and to have spacecraft-to-safe mode autonomous procedures in place in case of the extremely unlikely case of an ACS upset event. The effects of surface and jet winds can also be minimized by placing modern "furling/folding" or rotatable solar panels into a stowed, parallel-to-the-surface-normal position during close surface flybys, or utilizing separate "aerodynamic" daughter spacecraft with much lower surface area/mass ratios for all close-in work.

Confusion of guidance star trackers by sunlit dust particles flying past the spacecraft can be addressed using the next generation of star trackers implementing improved transient rejection algorithms. Potential damage from dust coverage and/or scouring of spacecraft surfaces can be mitigated by including closable shutters on all instruments, continual performance monitoring of optical instruments and solar panels, enforcing a "no remote sensing observations staring down a jet axis when within 10 km of the nucleus" policy, and avoiding flying spacecraft through outflowing jets.

Future missions can expect that significant scientific and engineering resources will be dedicated to near-comet mission planning, including allocating sufficient time (weeks to months) of time at ~100 km from the nucleus to produce global maps, and testing of parabolic flyby sweeps and close-in spacecraft hovering navigation in the highly asymmetric, spinning, multipole gravitational field of the comet's nucleus.

Finally, we provide a summary checklist of recommendations any s/c operating in the near nucleus zone of a comet needs to consider:

- In order to avoid actively outgassing regions and their nearby environs, detailed mapping of active surface regions will be required to allow adequate trajectory avoidance. Repetitive far-field (>500 km) approach imaging should give the rough location of any major jets.

- Gas sensors can detect important increases in total gas density and comet output ($Q_{gas}$) to independently verify the location of actively outgassing regions.

- At the closest approach distances, a spacecraft accelerometer with 0.5 ng sensitivity should be able to detect the accelerations due to outgassing at r < 3 AU, and due to major jets at r < 2 AU.

- The effects of cometary outgassing could be minimized by actively decreasing the spacecraft's surface area to mass ratio (e.g., by implementing rotatable solar panels or only using detachable, "aerodynamic" daughter spacecraft inside 100 km).

- Controllable, navigable solar radiation pressure forces dominate outside 4 AU, and can dominate for spacecraft-comet distances > 30 $R_{nuc}$ (~15 km) at 3 AU, > 100 $R_{nuc}$ (~50 km) at 2 AU, and 300 $R_{nuc}$ (150 km) at 1 AU.

- While cometary outgassing dominates cometary gravity for r < 2.5 AU, both forces are important to characterize and compensate for in order to properly navigate in the near-nucleus region.

- Near-comet operations involving optical instruments (in particular site selection for a landing and/or sample return, which require close passes (see Fig. 4) should be implemented if at all possible at relatively large heliocentric distances in order to minimize the deposited dust burdens on the s/c.

- Optical instruments should be equipped with front covers so as to protect optical surfaces from dust coverage when not operating, and continual performance monitoring of optical instruments and solar panels should be performed.

- Repetitive approach (>500 km) imaging should give timing of any regular flares or outbursts. In the extremely unlikely event of a massive comet calving/fragmentation event, contingency plans to remove the s/c to safe orbit distances of ~$10^3$ km from the comet center of mass and towards the Sun for a few days can be put in place.

- During the time period when the spacecraft is in the near-nucleus region, significant mission operations resources will need to be allocated for prox-ops navigation planning & implementation.

- The spacecraft star trackers will need to be able to handle up to $10^5$ discrete objects/image that are non-sidereal in conjunction with onboard IMUs/gyros in order to maintain accurate position knowledge for trajectories passing within 5 km of the nucleus surface.

- The main effect of all these issues is not expected to impact spacecraft health, but instead spacecraft position knowledge.

## 7. Acknowledgements

We would like to acknowledge the expert help of S.M. Byram and E. Gruen, in putting together the operations analysis presented in this paper, as well as the work of 3 anonymous reviewers. We would also like to acknowledge all the cometary spacecraft exploration pioneers who have come before us, leaders like Uwe Keller and Gerhard Schwehm and Andrea Accomazzo, and especially the leaders who are no longer with us like Michael A'Hearn, Michael Belton, and Fred Whipple. Of longer mention but no less important are the hundreds of pioneering scientists and engineers that have taken cometary science from its remote sensing infancy of flying sandbanks vs dirty snowballs in the 1950s to the precision rendezvous and sampling paradigm that exists today.